\documentclass[%
 reprint,
superscriptaddress,
bibnotes,
 amsmath,amssymb,
aps,
prb,
]{revtex4-1}
\usepackage{graphicx}% Include figure files
\usepackage{dcolumn}% Align table columns on decimal point
\usepackage{bm}% bold math

\begin{document}
\bibliographystyle{apsrev4-1}

\preprint{APS/123-QED}

%\title{Probing a molecular electronic and nuclear spin via a carbon nanotube quantum dot.}

\title{Landau-Zener tunneling of a single Tb$^{3+}$ magnetic moment allowing the electronic read-out of a nuclear spin}% Force line breaks with \\

\author{M. Urdampilleta}
\affiliation{
Institut N\'eel, CNRS et Universit\'e Joseph Fourier, BP 166, F-38042 Grenoble Cedex 9, France 
}
\author{S. Klyatskaya}
\affiliation{
Institute of Nanotechnology (INT), Karlsruhe Institute of Technology (KIT),  
76344 Eggenstein-Leopoldshafen, Germany 
}

\author{M. Ruben}
\affiliation{
Institute of Nanotechnology (INT), Karlsruhe Institute of Technology (KIT),  
76344 Eggenstein-Leopoldshafen, Germany 
}
\affiliation{
Institut de Physique et Chimie des Mat\'eriaux de Strasbourg (IPCMS), CNRS-Universit\'e de Strasbourg, 
67034 Strasbourg, France 
}
\author{W. Wernsdorfer}
\affiliation{
Institut N\'eel, CNRS et Universit\'e Joseph Fourier, BP 166, F-38042 Grenoble Cedex 9, France
}

\date{\today}% It is always \today, today,

\begin{abstract}
A multi-terminal device based on a carbon nanotube quantum dot was used at very low temperature to probe a single electronic and nuclear spin embedded in a bis-phthalocyanin Terbium (III) complex (TbPc$_2$). 
A spin-valve signature with large conductance jumps was found when
two molecules were strongly coupled to the nanotube. The application of a transverse field separated the magnetic signal of both molecules and enabled single-shot read-out of the Terbium nuclear spin.
The Landau-Zener (LZ) quantum tunneling probability was studied as a function of field sweep rate, establishing a good agreement with the LZ equation and yielding the tunnel splitting $\Delta$. It was found that $\Delta$ increased linearly as a function of the transverse field. 
These studies are an essential prerequisite for the coherent manipulation of a single nuclear spin in TbPc$_2$.
\begin{description}
\item[PACS numbers]
75.50.Xx, 81.07.Nb, 81.07.Ta, 85.75.-d
\end{description}
\end{abstract}

\pacs{Valid PACS appear here}% PACS, the Physics and Astronomy
\maketitle

%\tableofcontents

\section{Introduction}
%%%%%%%%%%%%%%%%%%%%%%%%%%%%%%%%%
The detection and manipulation of single localized spins is a rapidly developing field of nanoscience, enabling the creation of quantum bits and memories. Among the large variety of spin systems, electron spins confined in  quantum dots \cite{Loss1998,Imamoglu1999,Elzerman2004,Stotz2005,Koppens2006}, impurities,\cite{Steger2012} defects in semiconductors,\cite{Koehl2011}  and nitrogen-vacancy centers \cite{Dutt2007,Neumann2010,Robledo2011,Dreau2013} are promising candidates for spin-based quantum computation. However, control of the decoherence induced by the environment \cite{Huang2011} and control of the coupling between different spins are still experimental challenges. Single-molecule magnets (SMMs) offer an interesting alternative for the embedding of spins in a controllable and reproducible environment.\cite{Heersche2006,Bogani2008, Zyazin2010,Urdampilleta2011, Vincent2012, Marc2013}

A SMM consists of an inner magnetic core, containing in general several magnetic atoms, and an outer non-magnetic shell, which is chemically tailored through the addition of ligands.\cite{Sessoli1993,Christou2000} At low temperatures, the magnetic core of an SMM behaves like a single magnetic moment, which can attain spin ground states up to $S=83/2$.\cite{Ako2006} SMMs are characterized by a broad range of quantum effects, ranging from quantum tunneling of magnetization (QTM) \cite{Friedman1996, Thomas1996} to Berry-phase interference.\cite{Wernsdorfer1999} Their spin coherence time is expected to reach the microsecond range\cite{Schlegel2008, Ardavan2007}, and the perspective of entanglement between SMMs \cite{Timco2009} makes them good candidates for quantum spintronics \cite{Bogani2008} and quantum computations.\cite{Leuenberger2001}   
Among the large variety of known SMMs, rare earth based SMMs are of particular interest because the high intrinsic anisotropy of rare earth elements can lead to larger zero-field splittings than for transition metal based SMMs.\cite{Gatteschi2011} Furthermore, the strong hyperfine interaction of rare-earth elements leads to well separate resonant tunnel transitions for each nuclear spin state.\cite{Ishikawa2005} 
Because of the excellent intrinsic isolation of nuclear spins, they are promising candidates for spin quantum bits.\cite{Leuenberger2002}
% and those coupled to electronic moments are of particular interest for entanglement cite\{Simmons2011} and electronic read-out cite\{Pla2013} schemes.

\begin{figure}
{
\includegraphics[scale=0.98]{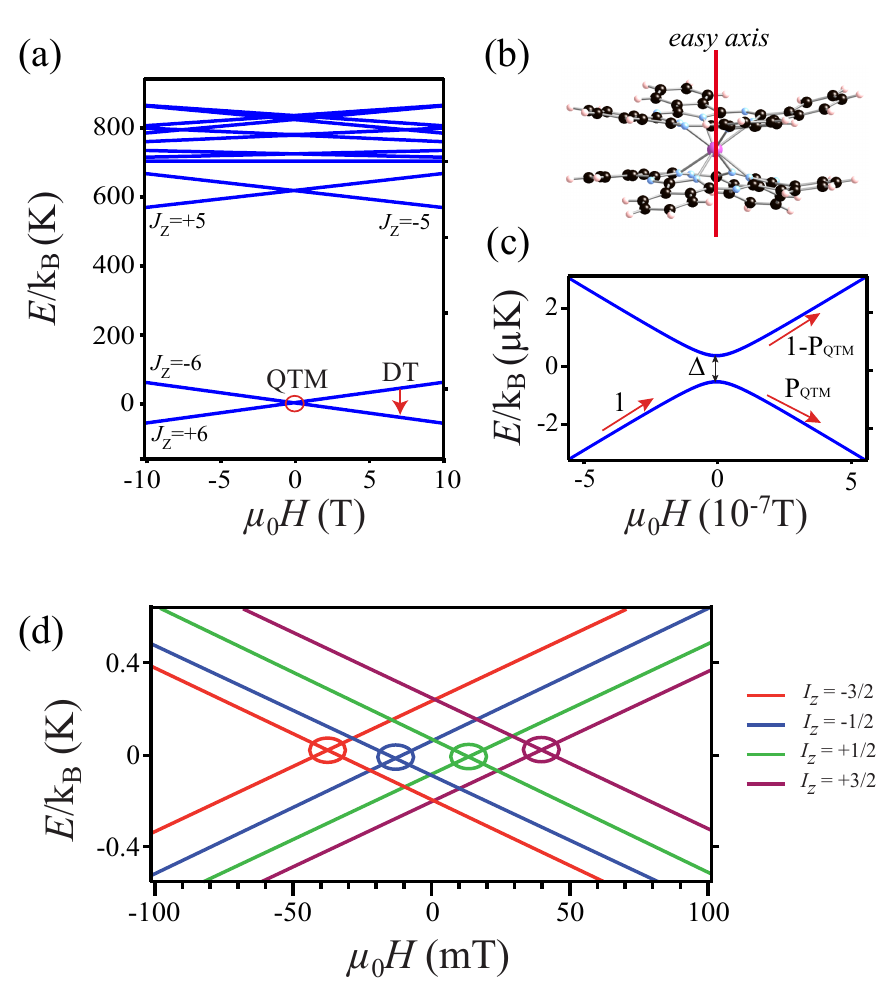}}
\caption{(Color online) (a) Zeeman diagram of the ground state $J=6$. $|J=6,J_z=\pm6\rangle$ is the low lying ground state doublet, which is separated by about 600 K from the first excited state $|J=6,J_z=\pm5\rangle$. The two mechanisms for moment reversal are quantum tunneling of the magnetization (QTM) occurring at zero magnetic field and the direct transition (DT) occurring at finite magnetic field. (b) Structure of the TbPc$_2$ complex. The black, blue, white and purple atoms correspond to C, N, H, and Tb, respectively . The magnetic easy axis and the hard plane are respectively orthogonal and parallel to the phtalocyanins. (c) Zoom on the avoided level crossing label by QTM in (a). $\Delta$ is the tunnel splitting, and $P_{QTM}$ is the tunnel probability. (d) Zeeman diagram of the ground doublet $|J_z=\pm6\rangle$ split by the strong hyperfine interaction of Tb. The color code illustrates the different nuclear spin states and the circled intersections are avoided level crossings. }
\label{fig1}
\end{figure}
 In this paper, we demonstrate the probing of single electronic and nuclear spins in rare earth based SMMs adsorbed onto the sidewall of a carbon nanotube. The present study focuses on the so-called Terbium "Double Decker" or bis-phthalocyanin Terbium (III) complex, referred to as TbPc$_{2}$ in the following, which was synthesized using the method reported in Ref.\cite{Klyatskaya2009}. 

%%%%%%%%%%%%%%%%%%%%%%%%%%%%%%%%%
\section{Spin Hamiltonian and Landau-Zener tunneling in T\lowercase{b}P\lowercase{c}$_{2}$}
%%%%%%%%%%%%%%%%%%%%%%%%%%%%%%%%%
The TbPc$_{2}$ complex is composed of two organic phtalocyanins (Pc$_{2}$) coordinating a single Terbium ion in a square antiprismatic geometry [Fig.~1(b)]. The 4f$^{8}$ electronic shell of  the Tb$^{3+}$ leads to a total magnetic moment $J=L+S=3+3$. In addition, the Tb ion has a nuclear spin $I=3/2$. The $\pi$-type electronic structure of Pc$_{2}$ permits coupling between the Tb and the environment.\cite{Rizzini2011, Komeda2011}  
In order to calculate the Zeeman diagram of TbPc$_2$ [Fig.~1(a)], we used the ligand field Hamiltonian expressed in the Stevens operator formalism:\cite{Stevens1952}
\begin{equation}
\mathcal{H}_{lf}=\alpha A^0_2O^0_2+\beta (A^0_4O^0_4+A^4_4O^4_4)+\gamma A^0_6O^0_6,
\end{equation}
and the Zeeman Hamiltonian:
\begin{equation}
\mathcal{H}_{Zeeman}= g_J\mu_0\mu_B\textbf{J}.\textbf{H},
\end{equation}
where the $O^k_q$ are the Stevens operators \cite{Stevens}, $A^k_q$ the ligand field parameters from Ref. \cite{Ishikawa2003}, $\alpha$, $\beta$ and $\gamma$ the Stevens constant tabulated in Ref. \cite{Abragam1970}, $g_J$ the gyromagnetic factor of Tb, $\textbf{J}$ the electronic magnetic moment operator of Tb, and $H$ the applied magnetic field.
Numerical diagonalization of $\mathcal{H}_{Zeeman}+\mathcal{H}_{lf}$ gives $|J=6,J_z=\pm6\rangle$ as a ground state doublet, isolated from the first excited state $|J=6,J_z=\pm5\rangle$, by about $600\,K$ [Fig.~1(a)]. Therefore, the Tb magnetic moment is aligned with the quantization axis (easy axis) within the low temperature and low magnetic field regime (T$<1\,$K, $\mu_0H<2\,$T) and is considered as an Ising-type spin in the following. 
The presence of a transverse anisotropy in the ligand field Hamiltonian ($A^4_4O^4_4$) couples the two states $|J_z=\pm6\rangle$, giving rise to an avoided level crossing with a tunnel splitting $\Delta \approx$~1~$\mu$K [Fig.~1(c)]. As a result, quantum tunneling of the magnetization (QTM) can occur in TbPc$_{2}$ when the magnetic field is swept through this avoided level crossing. 
The tunneling probability is given by the Landau-Zener equation:\cite{Wernsdorfer2005}
\begin{center}
\begin{equation}
\label{eqLZ}
\textit P_{QTM} = 1 - exp\left[  \dfrac{-\pi\Delta^2}{4\hbar g_J\mu_{\rm B}\mid J_z\mid\mu_{\rm 0}\dfrac{\mbox {d}H_z}{\mbox {d}t}}\right]
\end{equation}
\end{center}
where $d \textit H/\rm d \textit t$ is the magnetic field sweep rate and $J_z=\pm6$. 
When the electronic moment does not tunnel at the avoided level crossing, it ends up in the excited state and can relax to the ground state via a direct transition (DT). This process involves a spin-phonon interaction in which the Zeeman energy is released via a phonon. It occurs in the presence of a magnetic field typically larger than $0.2\,$T.\cite{Ishikawa2005}

 An important property of the Tb ion is the presence of a $I=3/2$ nuclear spin. The strong hyperfine interaction with the electronic  moment splits each ground state $\vert J=6,J_z=\pm6\rangle$ into four substates. We considered the following spin Hamiltonian containing the ligand field interaction, the Zeeman term, the hyperfine interaction and the quadrupolar term:
\begin{equation}
\label{eqH}
\mathcal{H} = \mathcal{H}_{LF}+\mathcal{H}_{Zeeman}+A\textbf{I}.\textbf{J}+P(I_{z}^{2}-\frac{1}{3}(\textbf I +1)\textbf I)
\end{equation}
where $A$ is the hyperfine constant, $P$ the quadrupolar constant and \textbf{I} the Tb nuclear magnetic moment operator. The corresponding Zeeman diagram of the ground multiplet is plotted in Fig.~1(d).
Among the sixteen level intersections, only the four encircled intersections correspond to avoided level crossings. Tunneling can take place at one of these avoided level crossings. It changes the electronic magnetic moment by $\Delta J_z= \pm 12$ but conserves the nuclear spin state. As a consequence, the measurement of the field position of QTM in TbPc$_{2}$ enables a direct read-out of the Tb nuclear spin state.

%%%%%%%%%%%%%%%%%%%%%%%%%%%%%%%%%
\section{Sample fabrication and Coulomb blockade}
%%%%%%%%%%%%%%%%%%%%%%%%%%%%%%%%%
QTM was studied in individual TbPc$_2$ complexes, laterally coupled to a carbon nanotube quantum dot. 
Catalyst islands of nanoporous alumina with Fe/Mo were first patterned on a SiO$_2$ surface with a metallic back-gate by optical lithography using LOR3A resist. The nanotubes were then grown in a Firstnano CVD oven at $750^\circ$C, whereby methane was used as a carbon source. Source, drain, and side gate electrodes were patterned by aligned electron-beam lithography, defining 200 nm long CNT junctions. The residual resist was removed by annealing the sample at 300$^\circ$C under Ar flow. The bis-phtalocyanin Tb(III) complexes were functionalized with one pyrene and six hexyl groups in order to improve their grafting onto the carbon nanotube \cite{Klyatskaya2009, Lopes2010} and prevent their crystallization. They were suspended in dichloromethane and drop-casted onto the nanotubes. Samples with large source-drain resistance ($>$100 k$\Omega$ at room temperature) were micro-bonded and measured in a dilution fridge with a base temperature of about 40 mK. The electronic temperature of the source and drain electrodes was estimated to be about 150 mK. The setup was equipped with two magnets, generating up to 1 T in two orthogonal directions. Magnetic moment reversal of the TbPc$_2$ was detected by means of conductance measurements through the carbon nanotube, as proposed previously.\cite{Bogani2008}
The differential conductance was measured as a function of side-gate and back-gate with an Adwin real-time acquisition system, programmed in a lock-in mode (100 $\mu$V and 67 Hz) at zero bias. Fig.~2 presents the differential conductance measurements of this four-terminals device, shown in the inset of Fig.~2. The conductance lines correspond to the Coulomb peaks (charge degeneracy state), and the zero-conductance region corresponds to Coulomb blockade. Both the side-gate and the back-gate are used to tune the chemical potential of the quantum dot, and to modify its coupling with the leads and the TbPc$_2$ molecules. 

\begin{figure}
{\label{fig2}
\includegraphics[scale=0.98]{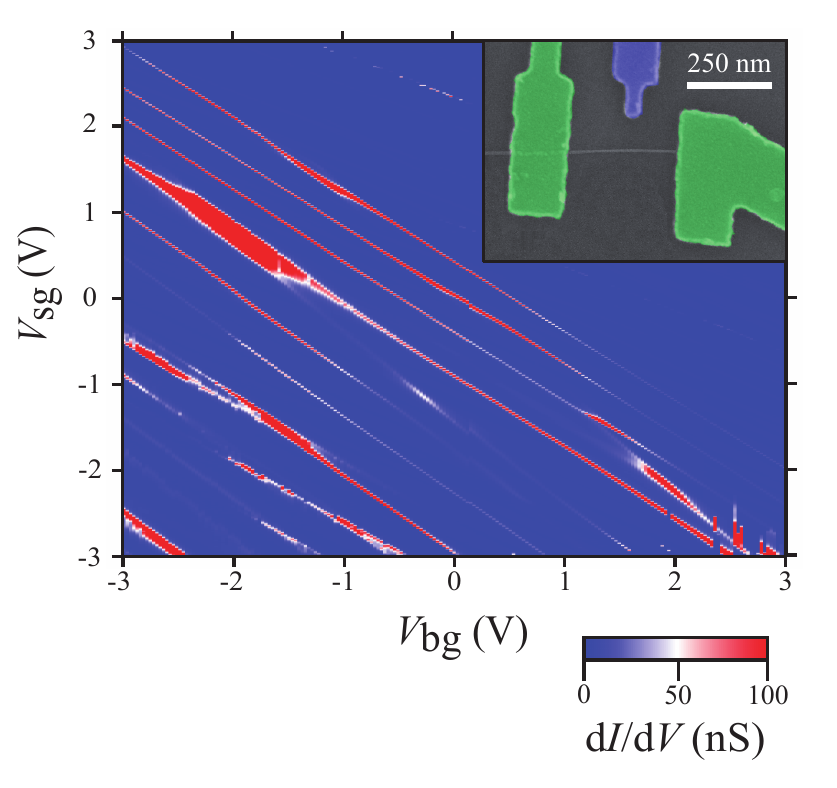}}
\caption{(Color online) Stability diagram of the device shown in the inset. The red lines are degeneracy points in the Coulomb blockade regime. Coupling between the leads and the quantum dot can be tuned by moving along one of these lines, by means of the back-gate and side-gate voltages ($V_{\rm bg}$ and $V_{\rm sg}$, respectively). Inset: Scanning electron micrograph of the studied device. The green electrodes correspond to the source and drain, and the blue electrode corresponds to the side gate.}
\end{figure}
%%%%%%%%%%%%%%%%%%%%%%%%%%%%%%%%%
\section{The spin-valve features}
%%%%%%%%%%%%%%%%%%%%%%%%%%%%%%%%%
We recently discovered a surprisingly strong interaction between the charge carriers of a carbon nanotube and TbPc$_{2}$ molecules grafted onto its sidewall.\cite{Urdampilleta2011} We found that this interaction gives rise to a spin-valve signature in the magneto-transport measurements: when the magnetic moment of two molecules are aligned in the parallel configuration the conductance is maximum while in the antiparallel configuration the conductance is minimum. Even so two theoretical studies proposed an interpretation of the underlying mechanism, \cite{Hong2013, Renani2013} more studies are needed to fully understand this effect.

For the present study, the conductance of the device in Fig.~2 was recorded as a function of the magnetic field for each non-zero conductance region. The spin-valve feature presented in Fig.~3(a) was observed at $V_{bg}=0\,$V and $V_{sg}=0\,$V. Along the corresponding degeneracy line, the signal vanished. One reason for this gate dependence could be that the coupling between the molecule and the nanotube, mediated by the $\pi$ type electrons of the Pc$_2$, was changed by the electric field. Indeed, it was shown by Lodi Rizzini et al. \cite{Rizzini2011} that the exchange coupling between the Tb and the environment is very sensitive to the oxidation state of the Pc$_2$.  This possible explanation has to be strengthened by further experiments and theoretical calculations.

This spin dependent effect is best visible when the two molecules have different coercive fields: in this case, each magnetic moment reversal induces a sharp change in the conductance of the device at a different field. 
In the present sample, one of these molecules (referred to as molecule A in the following) had a QTM probability close to one, inside the experimentally resolved sweep rate window, while the other molecule (molecule B) had a sweep rate dependent probability. It is not surprising that both molecules had a different tunnel splitting amplitude, since any deformation of the molecule (e.g. a slight twist between both Pc) generates a significant change in the longitudinal and transverse anisotropy.\cite{Sorace2011} 

For instance, for rather fast field sweep rates, A experienced QTM while B experienced a DT. The corresponding magneto-transport measurement are presented in Fig.~3(a). For slow field sweep rates, the magnetic moment of both molecules reversed by tunneling [Fig.~3(b)] and it was therefore difficult to distinguish between molecule A and B. However, because the molecules had their easy axes aligned in two different directions [Fig.~4(a)], with a constant transverse field $H_{\perp}$, orthogonal to the easy axis of molecule B, we could discriminate between the QTM positions of molecules A and B. Indeed, as depict in Fig.~4a, the longitudinal field felt by molecule A is $H_\perp Sin\theta+H_\parallel Cos\theta$ while the field felt by molecule B is simply $H_\parallel$, where $\theta$ is the angle between the easy axes of molecules A and B. More details concerning the magnetic field alignment are presented in the following section.

\begin{figure}
{\label{fig3}
\includegraphics[scale=0.98]{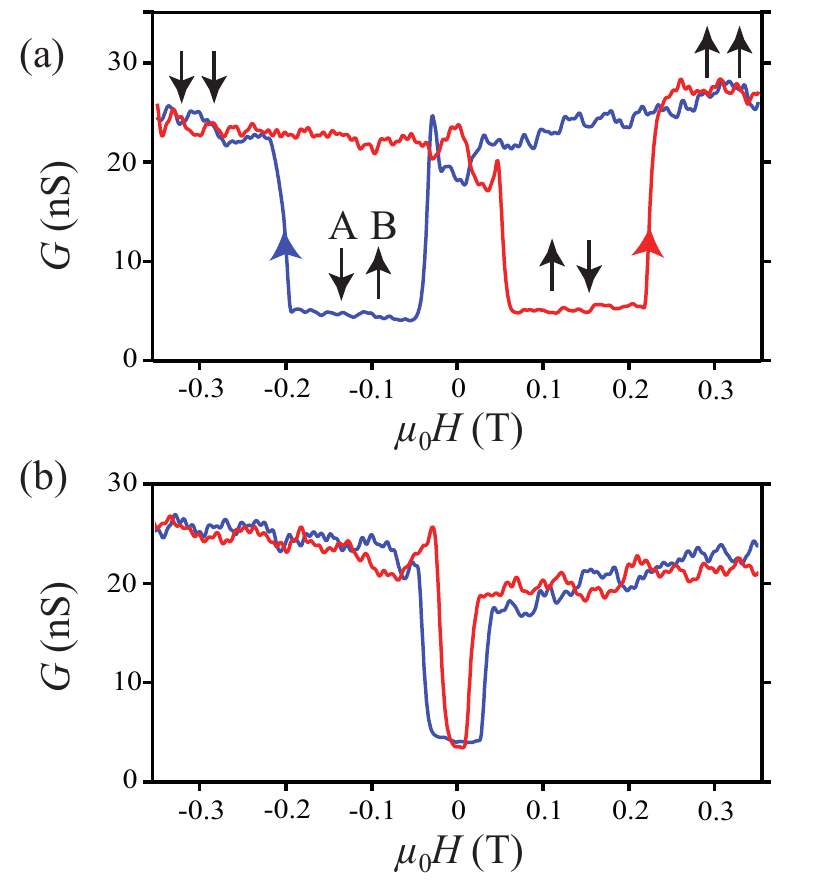}}
\caption{(Color online) (a) Magneto-conductance recorded at a sweep rate of $50\,$mT.s$^{-1}$. The blue curve corresponds to the measurements made from $-0.35\,$T up to $+0.35\,$T, and the red curve corresponds to the ones made from $+0.35\,$T down to $-0.35\,$T. Each of the sharp changes in conductance corresponds to the signature of a single TbPc$_{2}$. The reversal, which takes place under a low magnetic field, is due to QTM, whereas that which takes place under a higher magnetic field is due to DT. (b) Magneto-conductance recorded at a sweep rate of $20\,$mT.s$^{-1}$, the colour code is the same as for (a). The sharp conductance changes occur close to zero magnetic field and are the consequence of both molecules experiencing QTM.}
\end{figure}

%%%%%%%%%%%%%%%%%%%%%%%%%%%%%%%%%
%\section{Discrimination of QTM positions by application of a transverse field}
\section{Determination of the easy axis of magnetization}
%%%%%%%%%%%%%%%%%%%%%%%%%%%%%%%%%
In order to determine the easy axis of magnetization of molecule B, the spin-valve effect was measured as a function of the applied magnetic field angle. For each angle we measured the magnetoconductance for the trace (from $-1\,$T to $+1\,$T) and the retrace (from $+1\,$T to $-1\,$T).  A relative fast sweep rate of 50$\,$mT.s$^{-1}$ was used to have a high probability for a DT of molecule B. Fig.~4(b) shows the difference between retrace and trace for positive magnetic fields and the difference between trace and retrace for negative fields.
The red areas between $-70^\circ$ and $+70^\circ$ correspond to curves similar to Fig.~3(a), whereas the region between $+70^\circ$ and $+110^\circ$ correspond to a situation where the magnetic field is not strong enough to allow a DT [see Fig.~3(b) of Ref. \cite{Urdampilleta2011b}]. 
%The $0^\circ$  orientation was thus defined as the projection of the easy axis onto the applied magnetic field plane. The $90^\circ$ orientation is therefore lying in the hard axis of molecule B [Fig.~4(b)]. 
From this characterization, the easy axis of molecule B was found to be orthogonal to the nanotube axis
As a consequence, the transverse field was therefore applied along the nanotube axis.

\begin{figure} {\label{fig4} \includegraphics[scale=0.98]{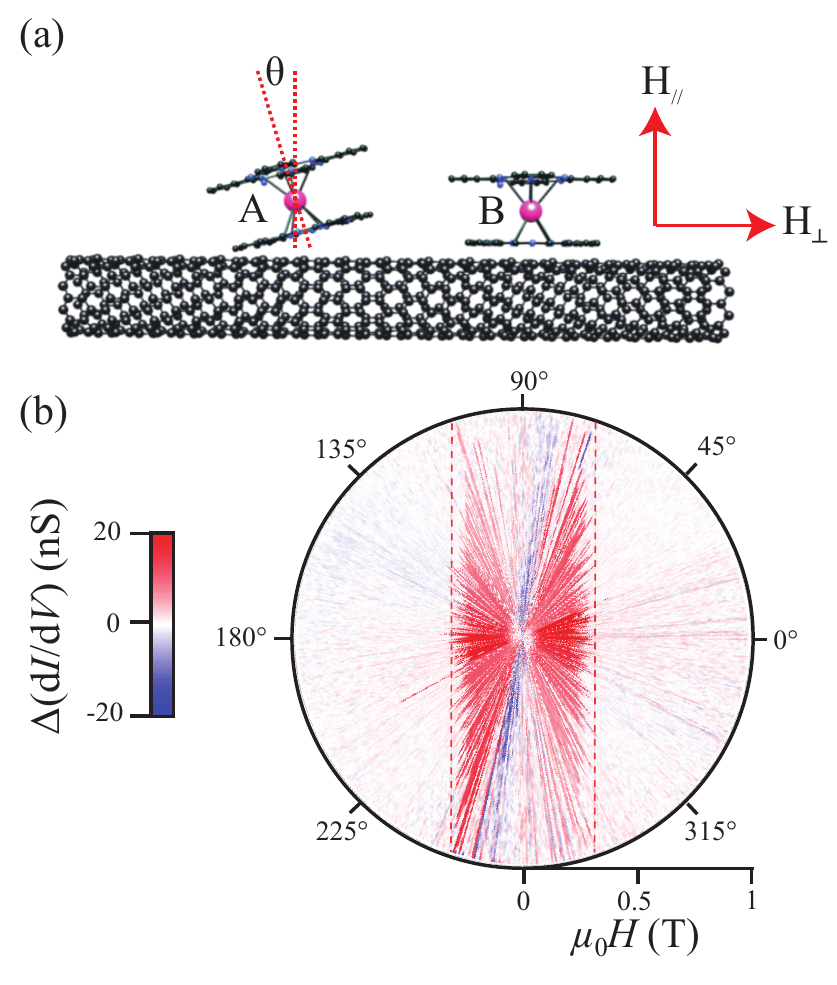}}
\caption{(Color online) (a) Schematic of the relative orientation of the two molecules on the nanotube. We define the longitudinal $H_{\parallel}$ and  transverse $H_{\perp}$ with respect to the molecule B which owns its easy axis perpendicular to the nanotube axis (see Fig.~4(a)), and $\theta$ the angle between the easy axis of B and A. (b) Anisotropy of the direct transition. The difference between the trace (from $-1\,$T to $+1\,$T) and the retrace (from $+1\,$T to $-1\,$T) is recorded as a function of angular orientation. The locus of the direct transition reversal is indicated by dashed lines. It is important to note that the reversal close to the $90^\circ$ orientation cannot occur because of the field limitations imposed by the magnetic field coils in the experimental setup.}
\end{figure}

\begin{figure} {\includegraphics[scale=0.98]{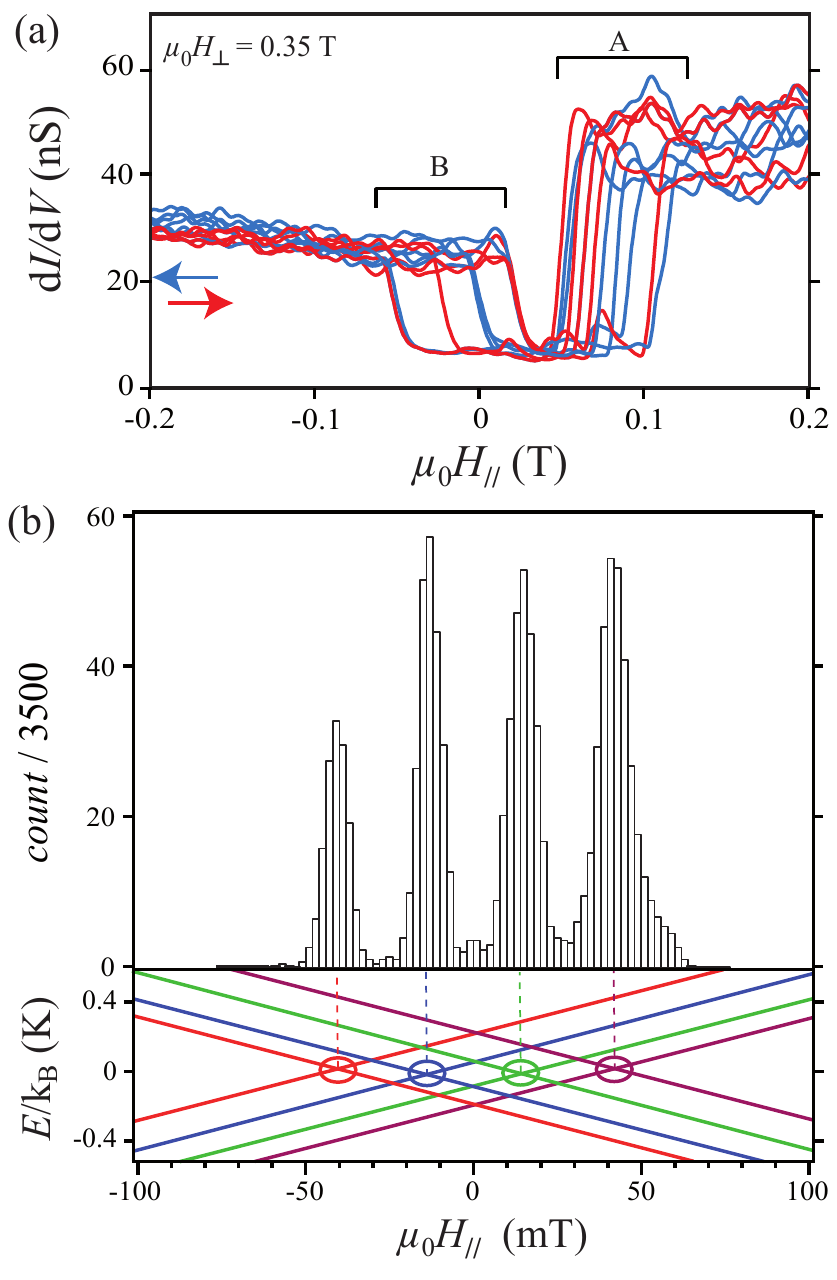}}
\caption{(Color online) (a) Magneto-conductance curves recorded under a $0.35\,$T transverse field, at a $100\,$mT.s$^{-1}$ sweep rate. The relative QTM position for molecule A and B are clearly split, the angle between their easy axes is then estimated around 15$^\circ$. (b) Histogram of the QTM position of molecule B, for 3500 consecutive traces recorded under the same conditions as those applied in (a). The position of those peaks corresponds to the avoided level-crossing shown on the Zeeman diagram of the ground doublet $J_z=\pm6$ split by the hyperfine coupling with the nuclear spin $I=3/2$.}
\label{fig5}
\end{figure}

%%%%%%%%%%%%%%%%%%%%%%%%%%%%%%%%%
\section{Electronic read-out of a single nuclear spin}
%%%%%%%%%%%%%%%%%%%%%%%%%%%%%%%%%
Figure 5a shows measurements recorded with a constant transverse field of $0.35\,$T and presenting only QTM features, the ones presenting DT being rejected. The conductance suddenly decreases, between the magnetic field values of $-50\,$mT and $+50\,$mT (QTM of B), and then increases in all cases above approximately $+50\,$mT (QTM of A).  This measurement was repeated 3500 times with a $100\,$mT.s$^{-1}$ sweep rate. Whenever DT failed to occur, the longitudinal position of the QTM in molecule B was stored in the histogram plotted in Fig.~5b: four peaks emerge with a FWHM of approximately 10 mT, and a mean peak-to-peak separation of 25 mT.

In order to explain these results, we compare the position of these peaks with the Zeeman diagram. Fig. 5(b) shows the very good correspondence between the four peaks and the avoided-level crossing of the Zeeman diagram. This diagram is slightly different from the one  presented in Fig.~1(c) since we took into account that the easy axis is not lying exactly in the plane ($H_{\parallel}$,$H_{\perp}$). As evident from the comparison between the histogram and the diagram, each of these peaks can be attributed to a particular nuclear spin state.

Similar results have recently been demonstrated at the single molecule level by Vincent et al. \cite{Vincent2012}, in a molecular transistor configuration. It is important to note that in the present case the FWHM is larger than in the case of Vincent et al. leading to a lower fidelity in the single-shot read-out measurement. One reason for this could be that the current tunneling through the carbone nanotube is interacting more strongly with the TbPc$_2$ molecules. 
Nevertheless, one advantage of the present device is the very large variation of conductance induced by the spin reversal ($200\%$ in the present case versus $1\%$ in the work of Vincent et al.\cite{Vincent2012}), which makes the measurement very easy since it does not require any specific filtering (physical or numerical).

In order to confirm that only one QTM position exists per nuclear spin state, we measured the tunneling probability as a function of the sweep rate and compared it with the Landau-Zener theory. 100 magneto-conductance measurements were recorded for a given sweep rate. The tunneling probability $P_{QTM}$ of molecule B can be obtained by $P_{QTM}=1-P_{DT}$, where $P_{DT}$ is the probability of a DT [Fig.~1(c)]. $P_{QTM}$ is plotted in Fig.~6 where the experimental data were fit with Eq. \ref{eqLZ}, from which a tunnel splitting of $\Delta=1.7\,\mu$K was extracted. As a consequence, the mono-exponential behavior clearly demonstrates that only the four circled level crossings in Fig.~1(d) are avoided level-crossings. This is in agreement with the work of Vincent et al.\cite{Vincent2012} but not with the measurement done on a single crystal of TbPc$_2$. The latter case presents QTM at all intersections of the diagram depicted in Fig.~1(d), see the work of Ishikawa et al. \cite{Ishikawa2005} In an assembly of molecules, coupled with weak dipole interactions, multi-spin tunnel effects\cite{WW_PRL02,WW_PRB05} might be responsible for this observation but further investigations are needed to better understand this issue. 
\begin{figure}
{\includegraphics[scale=0.98]{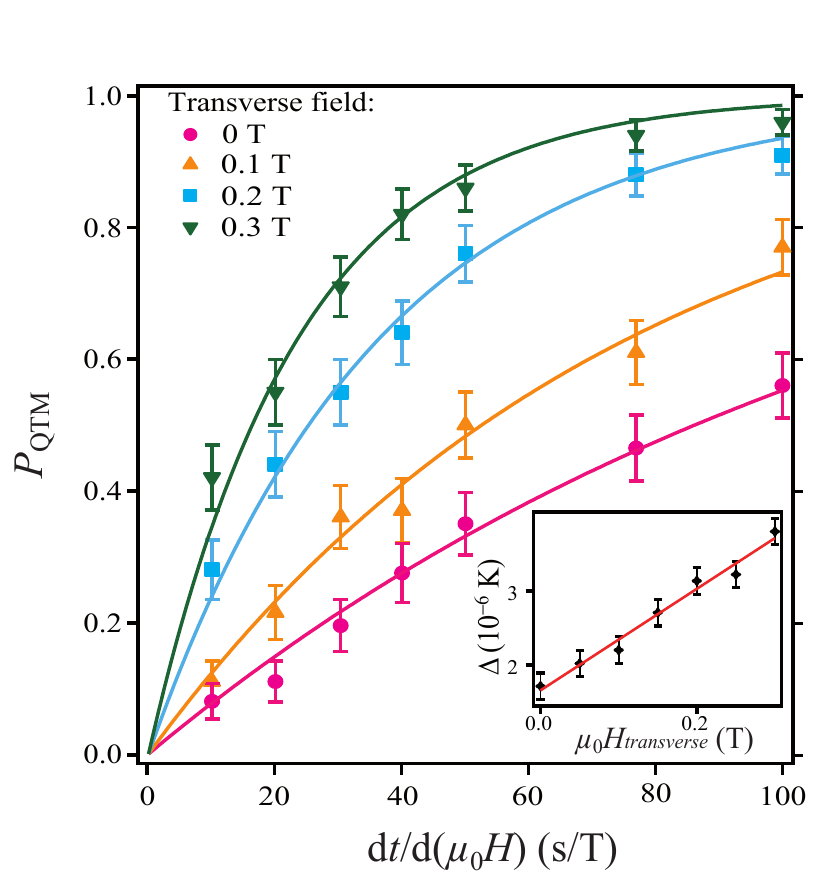}}
\caption{(Color online) QTM probability as a function of the inverse of the sweep rate, for different transverse fields. The experimental data are fitted by exponential curves, in accordance with equation (\ref{eqLZ}). The inset indicates the variation of tunnel splitting as a function of the transverse field, the red line is a guide for the eye.}
\label{fig6}
\end{figure}

%%%%%%%%%%%%%%%%%%%%%%%%%%%%%%%%%
\section{Landau-Zener tunneling in a single T\lowercase{b}P\lowercase{c}$_{2}$}
%%%%%%%%%%%%%%%%%%%%%%%%%%%%%%%%%
An applied transverse field tunes the tunnel splittings via the $H_\perp(J_++J_-)$ term of the Hamiltonian.
In order to study this effect on the different level crossings, we measured the tunneling probability for several constant transverse fields and field sweep rates. The symbols in Fig.~6  correspond to the experimental points and the continuous lines are least-square fits using Eq. (\ref{eqLZ}). The data agree very well with the Landau-Zener behavior, which suggests that no other measurable avoided-level crossings are induced by the application of a transverse field. The tunnel splitting amplitudes were extracted from the fits, and then plotted as a function of the transverse field (inset of Fig.~6). This behavior cannot be explained by using the parameters of Ishikawa et al. \cite{Ishikawa2005} Firstly, the ligand field might be different from the mean bulk value because the grafted molecules are probably slightly distorted and their anisotropy modified.\cite{Sorace2011} Secondly, the ligand field Hamiltonian does not predict a linear increase as observed for our measurements, which were confirmed on other molecular devices. This observation needs further experimental and theoretical investigations. In particular, we believe that, in the case of single molecules, the angular moment conservation has to be taken into account. When the latter is not conserved, the Landau-Zener equation is not valid.\cite{Marc2013b} Nevertheless, our studies showed that the hyperfine interaction is a robust feature allowing us to read the nuclear spin state regardless the deposition and measurement techniques. Moreover, the possibility of tuning the tunnel splitting is very convenient for experiences of coherent nuclear spin manipulation since the read-out mechanism needs the right value: not too small (no read-out) and not too large (DT possible, reducing the read-out fidelity). 

%%%%%%%%%%%%%%%%%%%%%%%%%%%%%%%%%
\section{Conclusion}
%%%%%%%%%%%%%%%%%%%%%%%%%%%%%%%%%
We quantitatively investigated quantum tunneling of magnetization at the single-molecule level and confirmed that a nanotube-based device is one of the best means for the detection of single magnetic moment reversal. Furthermore, we demonstrated that the nanotube device used for these measurements was able to read-out in a single-shot the state of an individual nuclear spin. The demonstrated tunability of the tunnel splitting, achieved by applying a transverse field, provides the possibility to adjust the quantum dynamics of such quantum spintronic devices.

%%%%%%%%%%%%%%%%%%%%%%%%%%%%%%%%%
\section{Acknowledgement}
%%%%%%%%%%%%%%%%%%%%%%%%%%%%%%%%%
 This  work  was  partially  supported  by  the  DFG  programs  SPP  1459  and  TRR  88  3Met,  
ANR-PNANO  project  MolNanoSpin  No  ANR-08-NANO-002,  ERC  Advanced  Grant  MolNanoSpin 
No  226558,  STEP  MolSpinQIP  and  the  Nanosciences  Foundation  of  Grenoble.  The samples  were manufactured  at  the  NANOFAB  facilities  of  the  N\'eel  Institute. We wish to  thank  B.  Fernandez,  G.  Julie, S. Dufresne, T. Crozes and T. Fournier for their assistance in manufacturing the nanotube device. We also wish to thank F. Balestro, J.-P. Cleuziou, D. Feinberg, M. Ganzhorn, N.-V. Nguyen, T.Novotny, and  R.  Vincent  for  fruitful  discussions,  and  E. Bonet, C.  Thirion, and R.  Piquerel for the software development. 
We also thank E. Eyraud, R. Haettel, C. Hoarau, D. Lepoittevin and V. Reita for their technical support.


\begin{thebibliography}{25}
\bibitem{Stevens}
$O^0_q$ are linear combination of even power of the $J_z$ operator up to the q$^{th}$ degree and $O^4_4=c(J^4_++J^4_-)$.
\end{thebibliography}
\end{document}